\documentclass{article}

\usepackage[final,nonatbib]{tackling_climate_workshop_style}

\usepackage[utf8]{inputenc} 
\usepackage[T1]{fontenc}    
\usepackage[hidelinks]{hyperref}       
\usepackage{url}            
\usepackage{booktabs}       
\usepackage{amsfonts}       
\usepackage{nicefrac}       
\usepackage{microtype}      
\usepackage{chemformula}
\usepackage[style=ieee,doi=false]{biblatex}
\usepackage{bbm}
\usepackage{todonotes}

\usepackage{booktabs}

\usepackage{tikz}
\usepackage{pgfplots}
\pgfplotsset{compat=1.17} 
\usetikzlibrary{arrows.meta}
\usetikzlibrary{backgrounds}
\usepgfplotslibrary{patchplots}
\usepgfplotslibrary{fillbetween}
\pgfplotsset{%
    layers/standard/.define layer set={%
        background,axis background,axis grid,axis ticks,axis lines,axis tick labels,pre main,main,axis descriptions,axis foreground%
    }{
        grid style={/pgfplots/on layer=axis grid},%
        tick style={/pgfplots/on layer=axis ticks},%
        axis line style={/pgfplots/on layer=axis lines},%
        label style={/pgfplots/on layer=axis descriptions},%
        legend style={/pgfplots/on layer=axis descriptions},%
        title style={/pgfplots/on layer=axis descriptions},%
        colorbar style={/pgfplots/on layer=axis descriptions},%
        ticklabel style={/pgfplots/on layer=axis tick labels},%
        axis background@ style={/pgfplots/on layer=axis background},%
        3d box foreground style={/pgfplots/on layer=axis foreground},%
    },
}
\usepackage{subcaption}

\usepackage{siunitx}
\usepackage{cleveref}
\usepackage{bm}

\usepackage{algorithm}
\usepackage{algorithmicx}
\usepackage[noend]{algpseudocode}

\title{A POMDP Model for Safe Geological Carbon Sequestration}

\author{
  Anthony Corso\\
  Department of Aeronautics and Astronautics\\
  Stanford University\\
  Stanford, CA \\
  \texttt{acorso@stanford.edu} \\
  \And
  Yizheng Wang\\
  Department of Earth and Planetary Sciences\\
  Stanford University\\
  Stanford, CA \\
  \texttt{yizhengw@stanford.edu} \\
\And
  Markus Zechner\\
  Department of Earth and Planetary Sciences \\
  Stanford University\\
  Stanford, CA \\
  \texttt{mzechner@stanford.edu} \\
\And
  Jef Caers\\
  Department of Earth and Planetary Sciences\\
  Stanford University\\
  Stanford, CA \\
  \texttt{jcaers@stanford.edu} \\
\And
  Mykel Kochenderfer\\
  Department of Aeronautics and Astronautics\\
  Stanford University\\
  Stanford, CA \\
  \texttt{mykel@stanford.edu} \\
}

\begin{document}

\maketitle

\begin{abstract}
Geological carbon capture and sequestration (CCS), where \ch{CO2} is stored in subsurface formations, is a promising and scalable approach for reducing global emissions.However, if done incorrectly, it may lead to earthquakes and leakage of \ch{CO2} back to the surface, harming both humans and the environment. These risks are exacerbated by the large amount of uncertainty in the structure of the storage formation. For these reasons, we propose that CCS operations be modeled as a partially observable Markov decision process (POMDP) and decisions be informed using automated planning algorithms. To this end, we develop a simplified model of CCS operations based on a 2D spillpoint analysis that retains many of the challenges and safety considerations of the real-world problem. We show how off-the-shelf POMDP solvers outperform expert baselines for safe CCS planning. This POMDP model can be used as a test bed to drive the development of novel decision-making algorithms for CCS operations.

\end{abstract}

\section{Introduction}
While global warming continues to pose an existential risk to humanity, the latest IPCC report~\cite{portner2022climate} suggests that reducing emissions by half before 2030 will secure a livable future. A wide variety of measures will be necessary to reduce greenhouse gas emissions, but in all significant reports, carbon capture and sequestration (CCS), where carbon is permanently removed from the atmosphere, accounts for about one-quarter to two-thirds of the cumulative emission reduction~\cite{cozzi2020world,benson2008carbon}. Geological carbon storage, where \ch{CO2} is stored into subsurface formations such as saline aquifers, is a promising technique due to its scalability and the recent increases in the price of \ch{CO2}~\cite{metz2005ipcc,wilberforce2019outlook, international2020ccus}, and will be the focus of this work.  

Even though CCS projects are similar to other, well-studied, subsurface problems such as groundwater or oil and gas applications, they involve additional and unsolved challenges. First, much less is known about saline aquifers compared to oil and gas reservoirs, resulting in significantly higher subsurface uncertainty. Second, the interaction of supercritical \ch{CO2}, brine, and rock results in complex physical and chemical processes such as trapping mechanisms, mineral precipitations, and geomechanics that must be accounted for. Lastly, injection of \ch{CO2} into geological formations comes with significant risks such as induced earthquakes, fractured cap rocks, and reactivation of faults that can lead to leakage, harming human life and the environment. Therefore, any decision made in the course of a CCS project needs to balance the trade-off between safety and the utilization of storage capacity to keep the project economically viable. 

CCS operations require a large number of sequential decisions to be made under uncertainty, such as selecting aquifers from a portfolio, choosing injection rates, defining well locations, selecting information gathering campaigns, and selecting monitoring strategies. The inherent subsurface uncertainty combined with complex physio-chemical processes results in a challenging decision problem for CCS operations that require a tight coupling of information gathering and acting. For these reasons, we propose to formulate CCS as a partially observable Markov decision process (POMDP) and rely on POMDP solvers~\cite{kochenderfer_wheeler_wray_2022} to inform the decision-making process.  

\textbf{Related work}: Quantifying subsurface uncertainty is especially important for successfully implementing safe and effective CCS operations and monitoring. Although considerable research has been devoted to developing better uncertainty quantification techniques~\cite{scheidt2018quantifying,lu2021accurate}, significantly less attention has been paid to determining how to act on that information. Prior work has focused on non-sequential optimization approaches~\cite{isebor2014derivative,yang2013niched}, or the use of sequential approaches in other subsurface domains such as oil/gas~\cite{de2020reinforcement} and groundwater~\cite{wang2022sequential}.

\begin{figure}[t]
     \centering
     \begin{subfigure}[t]{0.48\textwidth}
        \centering
         \input{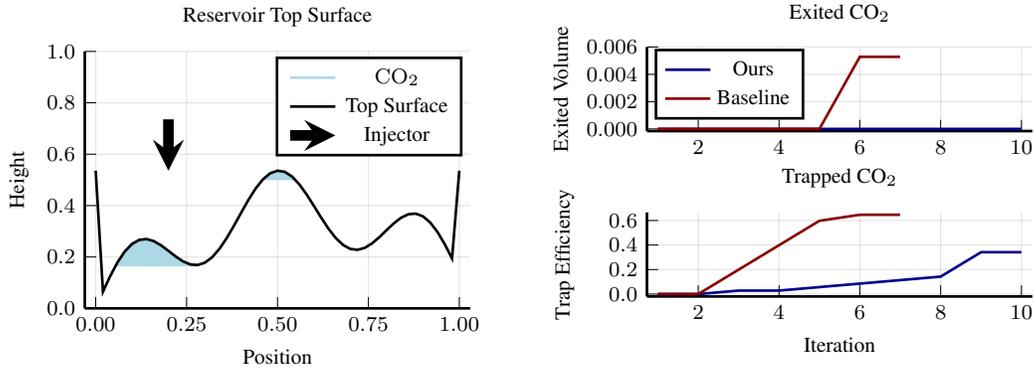}
         \caption{Spillpoint model for \ch{CO2} storage.}
         \label{fig:intro_left}
     \end{subfigure}%
     \begin{subfigure}[t]{0.48\textwidth}
        \centering

\begin{tikzpicture}[/tikz/background rectangle/.style={fill={rgb,1:red,1.0;green,1.0;blue,1.0}, draw opacity={1.0}}, show background rectangle]
\begin{axis}[point meta max={nan}, point meta min={nan}, legend cell align={left}, legend columns={1}, title={Exited CO$_2$}, title style={at={{(0.5,1)}}, anchor={south}, font={{\fontsize{8 pt}{10.4 pt}\selectfont}}, color={rgb,1:red,0.0;green,0.0;blue,0.0}, draw opacity={1.0}, rotate={0.0}}, legend style={color={rgb,1:red,0.0;green,0.0;blue,0.0}, draw opacity={1.0}, line width={1}, solid, fill={rgb,1:red,1.0;green,1.0;blue,1.0}, fill opacity={1.0}, text opacity={1.0}, font={{\fontsize{8 pt}{10.4 pt}\selectfont}}, text={rgb,1:red,0.0;green,0.0;blue,0.0}, cells={anchor={center}}, at={(0.02, 0.98)}, anchor={north west}}, axis background/.style={fill={rgb,1:red,1.0;green,1.0;blue,1.0}, opacity={1.0}}, anchor={north west}, xshift={1.0mm}, width=\textwidth, height={27mm}, scaled x ticks={false}, x tick style={color={rgb,1:red,0.0;green,0.0;blue,0.0}, opacity={1.0}}, x tick label style={color={rgb,1:red,0.0;green,0.0;blue,0.0}, opacity={1.0}, rotate={0}}, xlabel style={at={(ticklabel cs:0.5)}, anchor=near ticklabel, at={{(ticklabel cs:0.5)}}, anchor={near ticklabel}, font={{\fontsize{8 pt}{10.4 pt}\selectfont}}, color={rgb,1:red,0.0;green,0.0;blue,0.0}, draw opacity={1.0}, rotate={0.0}}, xmajorgrids={true}, xmin={0.73}, xmax={10.27}, xticklabels={{$2$,$4$,$6$,$8$,$10$}}, xtick={{2.0,4.0,6.0,8.0,10.0}}, xtick align={inside}, xticklabel style={font={{\fontsize{8 pt}{10.4 pt}\selectfont}}, color={rgb,1:red,0.0;green,0.0;blue,0.0}, draw opacity={1.0}, rotate={0.0}}, x grid style={color={rgb,1:red,0.0;green,0.0;blue,0.0}, draw opacity={0.1}, line width={0.5}, solid}, axis x line*={left}, x axis line style={color={rgb,1:red,0.0;green,0.0;blue,0.0}, draw opacity={1.0}, line width={1}, solid}, scaled y ticks={false}, ylabel={Exited Volume}, y tick style={color={rgb,1:red,0.0;green,0.0;blue,0.0}, opacity={1.0}}, y tick label style={color={rgb,1:red,0.0;green,0.0;blue,0.0}, opacity={1.0}, rotate={0}}, ylabel style={at={(ticklabel cs:0.5)}, anchor=near ticklabel, at={{(ticklabel cs:0.5)}}, anchor={near ticklabel}, font={{\fontsize{8 pt}{10.4 pt}\selectfont}}, color={rgb,1:red,0.0;green,0.0;blue,0.0}, draw opacity={1.0}, rotate={0.0}}, ymajorgrids={true}, ymin={-0.0001584658651284984}, ymax={0.006}, yticklabels={{$0.000$,$0.002$,$0.004$,$0.006$}}, ytick={{0.0,0.002,0.004,0.006}}, ytick align={inside}, yticklabel style={font={{\fontsize{8 pt}{10.4 pt}\selectfont}}, color={rgb,1:red,0.0;green,0.0;blue,0.0}, draw opacity={1.0}, rotate={0.0}}, y grid style={color={rgb,1:red,0.0;green,0.0;blue,0.0}, draw opacity={0.1}, line width={0.5}, solid}, axis y line*={left}, y axis line style={color={rgb,1:red,0.0;green,0.0;blue,0.0}, draw opacity={1.0}, line width={1}, solid}, colorbar={false}]
    \addplot[color={rgb,1:red,0.0;green,0.0;blue,0.5451}, name path={78cf695d-ddb1-4299-a461-523ed6009337}, draw opacity={1.0}, line width={1}, solid]
        table[row sep={\\}]
        {
            \\
            1.0  0.0  \\
            2.0  0.0  \\
            3.0  0.0  \\
            4.0  0.0  \\
            5.0  0.0  \\
            6.0  0.0  \\
            7.0  0.0  \\
            8.0  0.0  \\
            9.0  0.0  \\
            10.0  0.0  \\
        }
        ;
    \addlegendentry {Ours}
    \addplot[color={rgb,1:red,0.5451;green,0.0;blue,0.0}, name path={1931aa9a-7c4c-4c8e-b186-e6dba1e48fbe}, draw opacity={1.0}, line width={1}, solid]
        table[row sep={\\}]
        {
            \\
            1.0  0.0  \\
            2.0  0.0  \\
            3.0  0.0  \\
            4.0  0.0  \\
            5.0  0.0  \\
            6.0  0.00528219550428328  \\
            7.0  0.00528219550428328  \\
        }
        ;
    \addlegendentry {Baseline}
\end{axis}
\begin{axis}[point meta max={nan}, point meta min={nan}, legend cell align={left}, legend columns={1}, title={Trapped CO$_2$}, title style={at={{(0.5,1)}}, anchor={south}, font={{\fontsize{8 pt}{10.4 pt}\selectfont}}, color={rgb,1:red,0.0;green,0.0;blue,0.0}, draw opacity={1.0}, rotate={0.0}}, legend style={color={rgb,1:red,0.0;green,0.0;blue,0.0}, draw opacity={1.0}, line width={1}, solid, fill={rgb,1:red,1.0;green,1.0;blue,1.0}, fill opacity={1.0}, text opacity={1.0}, font={{\fontsize{8 pt}{10.4 pt}\selectfont}}, text={rgb,1:red,0.0;green,0.0;blue,0.0}, cells={anchor={center}}, at={(0.02, 0.98)}, anchor={north west}}, axis background/.style={fill={rgb,1:red,1.0;green,1.0;blue,1.0}, opacity={1.0}}, anchor={north west}, xshift={1.0mm}, yshift={-22.0mm}, width=\textwidth, height={27mm}, scaled x ticks={false}, xlabel={Iteration}, x tick style={color={rgb,1:red,0.0;green,0.0;blue,0.0}, opacity={1.0}}, x tick label style={color={rgb,1:red,0.0;green,0.0;blue,0.0}, opacity={1.0}, rotate={0}}, xlabel style={at={(ticklabel cs:0.5)}, anchor=near ticklabel, at={{(ticklabel cs:0.5)}}, anchor={near ticklabel}, font={{\fontsize{8 pt}{10.4 pt}\selectfont}}, color={rgb,1:red,0.0;green,0.0;blue,0.0}, draw opacity={1.0}, rotate={0.0}}, xmajorgrids={true}, xmin={0.73}, xmax={10.27}, xticklabels={{$2$,$4$,$6$,$8$,$10$}}, xtick={{2.0,4.0,6.0,8.0,10.0}}, xtick align={inside}, xticklabel style={font={{\fontsize{8 pt}{10.4 pt}\selectfont}}, color={rgb,1:red,0.0;green,0.0;blue,0.0}, draw opacity={1.0}, rotate={0.0}}, x grid style={color={rgb,1:red,0.0;green,0.0;blue,0.0}, draw opacity={0.1}, line width={0.5}, solid}, axis x line*={left}, x axis line style={color={rgb,1:red,0.0;green,0.0;blue,0.0}, draw opacity={1.0}, line width={1}, solid}, scaled y ticks={false}, ylabel={Trap Efficiency}, y tick style={color={rgb,1:red,0.0;green,0.0;blue,0.0}, opacity={1.0}}, y tick label style={color={rgb,1:red,0.0;green,0.0;blue,0.0}, opacity={1.0}, rotate={0}}, ylabel style={yshift=2.2mm, at={(ticklabel cs:0.5)}, anchor=near ticklabel, at={{(ticklabel cs:0.5)}}, anchor={near ticklabel}, font={{\fontsize{8 pt}{10.4 pt}\selectfont}}, color={rgb,1:red,0.0;green,0.0;blue,0.0}, draw opacity={1.0}, rotate={0.0}}, ymajorgrids={true}, ymin={-0.01939806225662969}, ymax={0.6660001374776193}, yticklabels={{$0.0$,$0.2$,$0.4$,$0.6$}}, ytick={{0.0,0.2,0.4,0.6}}, ytick align={inside}, yticklabel style={font={{\fontsize{8 pt}{10.4 pt}\selectfont}}, color={rgb,1:red,0.0;green,0.0;blue,0.0}, draw opacity={1.0}, rotate={0.0}}, y grid style={color={rgb,1:red,0.0;green,0.0;blue,0.0}, draw opacity={0.1}, line width={0.5}, solid}, axis y line*={left}, y axis line style={color={rgb,1:red,0.0;green,0.0;blue,0.0}, draw opacity={1.0}, line width={1}, solid}, colorbar={false}]
    \addplot[color={rgb,1:red,0.0;green,0.0;blue,0.5451}, name path={8b65020d-905b-4924-b2d6-0a1f4d669206}, draw opacity={1.0}, line width={1}, solid]
        table[row sep={\\}]
        {
            \\
            1.0  0.0  \\
            2.0  0.0  \\
            3.0  0.028462348786516836  \\
            4.0  0.028462348786516836  \\
            5.0  0.05692469757303367  \\
            6.0  0.08538704635955051  \\
            7.0  0.11384939514606734  \\
            8.0  0.1423117439325842  \\
            9.0  0.34154818543820203  \\
            10.0  0.34154818543820203  \\
        }
        ;
    \addplot[color={rgb,1:red,0.5451;green,0.0;blue,0.0}, name path={1ddf1989-9d9f-45c6-a977-da343e2fa8b0}, draw opacity={1.0}, line width={1}, solid]
        table[row sep={\\}]
        {
            \\
            1.0  0.0  \\
            2.0  0.0  \\
            3.0  0.19923644150561787  \\
            4.0  0.39847288301123573  \\
            5.0  0.5977093245168537  \\
            6.0  0.6466020752209897  \\
            7.0  0.6466020752209897  \\
        }
        ;
\end{axis}
\end{tikzpicture}%
         \caption{Performance metrics for a no-uncertainty baseline against POMCPOW (ours).}
         \label{fig:intro_right}
     \end{subfigure}
    \caption{A simplified model for CCS operations ammenable to automated decision making.}
    \label{fig:intro}
\end{figure}

In this work we introduce a POMDP model of CCS operations based on simplified physics in order to spur the development of more advanced solution techniques that can produce safe and reliable plans. We first describe our simplified model, which is based on spillpoint analysis, but retains many of the challenges associated with the full-scale CCS problem (\cref{fig:intro_left}). We then compare a variety of rule-based plans against state-of-the-art POMDP solvers and show how current off the shelf approaches can improve upon expert plans (\cref{fig:intro_right}). The spillpoint CCS model is implemented in the open source POMDPs.jl\footnote{\url{https://juliapomdp.github.io/POMDPs.jl/latest/}} framework. 

\section{Approach}

\paragraph{POMDP Background}
A partially observable Markov decision processes (POMDP)~\cite{kochenderfer_wheeler_wray_2022} is a model for sequential decision making where an agent takes actions over time to maximize the accumulation of a reward. A POMDP is defined by the tuple ($S$, $A$, $T$, $R$, $O$, $Z$, $\gamma$) where $S$ is the set of states, $A$ is the set of actions, $T$ is a transition model, $R$ is the reward function, $O$ is the observation space, $Z$ is the observation model, and $\gamma$ is the discount factor. For complex dynamical systems such as CCS, the transition function is implemented as a simulator from which transitions can be drawn according to the distribution $s^\prime \sim T(\cdot \mid s, a)$ defined by the action $a \in A$ taken in state $s \in S$. The agent is uncertain about the current state and therefore maintains a distribution over states called a belief $b(s)$. The initial belief $b_0(s)$ is updated at each timestep using observations $o \in O$ sampled with likelihood $Z(o \mid s, a, s^\prime)$. A policy $\pi$ is a function that maps beliefs to actions such that $a=\pi(b)$. POMDP solvers seek to approximate the optimal policy $\pi^*$ that maximizes the expected discounted sum of future rewards.


\paragraph{Challenges of Solving a CCS POMDP}
CCS can be formulated as a POMDP where the states represent the reservoir structure and the properties of the multi-phase fluid stored therein. The actions are the placement of sensors,  the drilling of \ch{CO2} injectors, and the setting of the rate of injection. The observations can come from many modalities including direct observation of the geology through drilling, as well as indirect methods such as seismic monitoring. The transition function is the defined by the complex fluid dynamics that govern the injection and evolution of the \ch{CO2} plume.

Optimal planning for CCS operations is challenging for several reasons: the state, action and observation spaces are continuous and high dimensional, the transition function is computationally intensive, requiring hours for a single transition, and since CCS operations are safety-critical, the plan must have a high degree of reliability. To encourage the development of algorithms that can solve planning problems of this scale, we developed a simplified model of CCS operations that retains many of these challenges, while requiring far less computational effort. 

\paragraph{A Spillpoint Model POMDP for CCS}
Buoyancy is one of the dominant physical forces that govern the dynamics of \ch{CO2} injection~\cite{MOLLNILSEN201533}. Spillpoint models, where only the effects of buoyancy are used to compute the dynamics of injected \ch{CO2}, are therefore a simple, but realistic model and have previously been used to estimate the trapping capacity of a reservoir given the top surface geometry~\cite{MOLLNILSEN201533}. In this work, we design a CCS POMDP using a 2D spillpoint model as the transition function to approximate the \ch{CO2} dynamics with low computational effort. 

In our model, the reservoir is defined by a top surface as shown in \cref{fig:intro_left}, and a porosity $\rho$ which defines the fraction of the volume available to store \ch{CO2}. Given an injection location and volume, the model determines the amount of \ch{CO2} trapped in each region of the reservoir. To compute \ch{CO2} saturation, the reservoir is automatically segmented into \emph{spill regions}, where \ch{CO2} is trapped due to buoyant forces, and \emph{spillpoints}, which are connections between spill regions, across which \ch{CO2} will migrate. For example in \cref{fig:intro_left}, \ch{CO2} is injected in the left spill region until the trapped volume exceeds the spill region volume, and it begins to fill the central spill region. If the spillpoint connects to a fault, then any \ch{CO2} that migrates through that spillpoint is considered to have exited the reservoir or \emph{leaked}. We model a fault on the left and right boundary of the reservoir domain. For additional details of the spillpoint algorithm, see \cref{sec:spillpointalgorithm}.

To construct a diverse set of reservoirs, we model the top surface geometry by the function 
\begin{equation}h = h_{\rm lobe} \sin(5 \pi x) + h_{\rm center}\sin(\pi x)+ h_{\rm elev} x
\end{equation}
where $x$ is the lateral position. The outer lobe height $h_{\rm lobe}$, central lobe height $h_{\rm center}$, and elevation slope $h_{\rm elev}$ determine the size and relationship between spill regions. The initial belief is implicitly defined by distributions over top surface parameters $h_{\rm lobe}$, $h_{\rm center}$, $h_{\rm elev}$, and the porosity $\rho$.

The rest of the POMDP is defined as follows. The state is defined by the top surface height of the reservoir, the porosity, the injector location, and the amount of trapped and exited \ch{CO2}. The actions include the position at which to drill the injector, the rate of \ch{CO2} injection, the ability to make noisy observations of the trapped \ch{CO2}, and the option to end CCS operations. The action to drill involves specifying a drill location and returns the top surface height at that location. The action to observe involves specifying sensor locations which return a measurement of the top surface height and \ch{CO2} thickness at that location, with noise distributed normally. If no \ch{CO2} is present, then \num{0} is returned for both values, providing no additional information on the top surface geometry. Additionally, we model a sensor on each of the boundaries of the domain that can noiselessly detect if \ch{CO2} is leaking, and is included in the observation at each transition, regardless of the action taken.

The reward function includes the competing objectives of storing \ch{CO2} while avoiding leakage. It has four components including trapped \ch{CO2} volume, exited \ch{CO2} volume, observation costs, and an indicator of \ch{CO2} leakage. Specifically, the reward is
\begin{equation}
R(s, a, s^\prime) = \lambda_{\textrm{leak}}\mathbbm{1}_{\{\Delta V_{\rm exited}>0\}} + \lambda_{\rm exited}\Delta V_{\rm exited} + \lambda_{\rm trapped}\Delta V_{\rm trapped} + \lambda_{\rm observation} N_{\rm obs} \label{eq:reward} \end{equation}
where $\mathbbm{1}_{\{\cdot\}}$ is the indicator function and $V_{\rm exited}$ and $V_{\rm exited}$ are, respectively, the changes in exited and trapped \ch{CO2} between states $s$ and $s^\prime$, and $N_{\rm obs}$ is the number of observation locations.

\section{Experiments}
We solve the spillpoint CCS POMDP with four different approaches. The first approach (\textbf{Random}) is a random policy where the actions are selected randomly until CCS operations are stopped by chance or until leakage is detected, at which point operations are halted. The second approach (\textbf{Best Guess MDP}) we do not consider uncertainty in the state, and instead start by selecting a top surface geometry that matches a single top surface measurement and treating it as the ground truth. Then Monte Carlo tree search is used to plan an optimal injection strategy. The third approach (\textbf{Fixed Schedule}) is an expert policy with a fixed observation schedule. The action to observer is used every three timesteps, incorporating each observation into a belief update. Injection continues as long as there is no possibility of leakage according to the current belief. The last two approaches include solving the formulation with an off-the-shelf POMDP solver called POMCPOW~\cite{sunberg2018online}. The first version of POMCPOW (\textbf{POMCPOW (Basic)}) uses a basic particle filter as the belief update strategy while the second version (\textbf{POMCPOW (SIR)}) uses the more advanced sequential importance resampling (SIR) belief updating approach (see \cref{sec:beliefupdating} for details).

To evaluate the various approaches, we randomly sampled \num{10} realizations of the reservoir geometry (as the ground truth), and ran each algorithm on this test set, recording the following \num{4} metrics. The first metric is the sum of discounted rewards (\textbf{Return}) defined in equation \ref{eq:reward}, which includes terms associated with total trapped \ch{CO2} as well as safety considerations. Additionally, we consider metrics associated with each sub component of the reward. These include the average number of observations per trajectory (\textbf{Observations}), the fraction of realizations that had leakage (\textbf{Leak Fraction}), and the fraction of trapped \ch{CO2} compared to the maximum possible trapped volume (\textbf{Trap Efficiency}). All hyperparameters are given in \cref{sec:hyperparameters}.

The results of these experiments are shown in \cref{tab:results}. The highest return and smallest amount of leakage is achieved by POMCPOW with SIR. This algorithm behaved conservatively, however, and rarely used the \emph{observe} action to gather more information, and therefore only reached an average of 27\% trap efficiency. Ignoring uncertainty lead to high trap efficiency but caused a significant chance of leakage at 50\%. The fixed observation schedule had an even higher trap efficiency (82\%) but still leaked in 10\% of cases. Using a naive belief updating procedure (particle filter) caused POMCPOW to become overconfident in its belief and therefore lead two cases with significant leakage. We therefore conclude that automated decision making systems can be helpful for safe CCS operations.

\begin{table*}
    \caption{Comparison of CCS policies\label{tab:results}}
    \centering
    \begin{tabular}{@{}lcccc@{}} 
    \toprule
    Method & Return & Observations & Leak Fraction & Trap Efficiency \\
    \midrule
    Random & $-104.38 \pm 320.41$  & $4.60 \pm 6.43$  & $0.10$  & $0.08 \pm 0.07$ \\
    Best Guess MDP & $-516.03 \pm 549.10$  & $0.00 \pm 0.00$  & $0.50$  & $0.70 \pm 0.27$ \\
    Fixed Schedule & $-97.91 \pm 316.28$  & $42.70 \pm 24.57$  & $0.10$  & $\bm{0.82 \pm 0.17}$ \\
    POMCPOW (Basic) & $-2811.32 \pm 6009.65$  & $1.20 \pm 1.55$  & $0.20$  & $0.58 \pm 0.30$ \\
    POMCPOW (SIR) & $\bm{2.24 \pm 4.04}$  & $3.00 \pm 3.46$  & $\bm{0.00}$  & $0.27 \pm 0.32$ \\
    \bottomrule
    \end{tabular}
\end{table*}

\section{Conclusion}
To address the upcoming challenges of safe geological carbon storage, we proposed to model CCS as a POMDP and use automated planning algorithms to design safe operational plans. We developed a simplified CCS formulation based on spillpoint analysis that eases the computational burden of exploring automated decision making for CCS and demonstrated the performance of several existing algorithms and baselines. We hope this model can be used to drive the development of new algorithms that support real-world CCS operations.

\begin{ack}
\end{ack}

\printbibliography

\appendix

\section{Spillpoint Algorithm}
\label{sec:spillpointalgorithm}
Spillpoint analysis~\cite{MOLLNILSEN201533} allows for the calculation of \ch{CO2} saturation from the geometry of the reservoir and total amount of \ch{CO2} injected. The procedure takes as input the top surface geometry $h(x)$, the spill region $s_{r}$, and the total injected \ch{CO2} $V_{\rm inject}$ and returns the \ch{CO2} saturation at each position, the trapped volume and the exited volume. Details of the algorithm can be found in \cref{alg:spillpoint}. The \ch{CO2} saturation is represented a set of polyhedra that indicate the 2D area that is occupied by \ch{CO2}. The polyhedra are computed by first identifying the spill region responsible for trapping (we followed the procedure outlined by \citeauthor{MOLLNILSEN201533}), and then determining the depth of the \ch{CO2} plume. The depth is found by using an optimizer to minimize the difference between the injected volume and the volume of of the polyhedra for a given depth. This optimization procedure is the most expensive part of the computation. 

\begin{algorithm}
\caption{Algorithm for determining \ch{CO2} depth using spillpoint analysis for mesh $m$ and volume $V$} \label{alg:spillpoint}
\begin{algorithmic}[1]
    \Function{inject}{$h$, $s_r$, $V_{\rm inject}$}
    \State $V_{\rm trap} = 0$
    \State $SRs = \emptyset$ \Comment{Spill regions that are full}
    \State $Sat = \emptyset$ \Comment{Polyhedra representing saturated volume}
    \While{$V_{\rm trap} < V$} \Comment{While there is remaining \ch{CO2} volume}
        \State $V_{\rm remain} = V_{\rm inject} - V_{\rm trap}$ \Comment{Compute remaining \ch{CO2}}
        \State $V_{\rm sr} \gets \textsc{SpillRegionVolume}(s_r)$\Comment{Max storage of spill region}
        \If{$V_{\rm sr} > V_{\rm remain}$}
            \State $Sat \gets Sat \bigcup \textsc{GetPolys}(h, s_r, V_{\rm remain})$
            \State \textbf{break}
        \Else
            \State $Sat \gets Sat \bigcup \textsc{GetPolys}(h, s_r, V_{\rm sr})$
            \State $V_{\rm trap} \gets V_{\rm trap} + V_{\rm sr}$ \Comment{Count the new trapped \ch{CO2}}
            \State $SRs \gets SRs \bigcup s_r$ \Comment{Record full spill region}
            \State $sr \gets \textsc{UphillSpillRegion}(h, s_r)$ \Comment{Get the connecting spill region} 
            \If{$s_r \in \textsc{Faults}(h)$} \Comment{If next spill region is a fault}
                \State $V_{\rm exit} \gets V_{\rm remain}$
                \State \textbf{break}
            \EndIf
            \If{$s_r \in SRs$} \Comment{Adjacent spill regions are full}
                \State $h \gets \textsc{MergeSpillRegions}(h, s_r)$ \Comment{Merge into larger region}
                \State $V_{\rm trap} \gets 0$
                \State $SRs = \emptyset$
                \State $Sat = \emptyset$
            \EndIf
        \EndIf
    \EndWhile
    \State \Return{$Sat, V_{\rm trap}, V_{\rm exited}$}
    \EndFunction
\end{algorithmic}
\end{algorithm}

\section{Belief Updating}
\label{sec:beliefupdating}
After taking action $a$, the prior belief $b(s)$ is updated to the posterior belief $b(s^\prime)$ with the observation $o$ according to Bayes' rule~\cite{kochenderfer_wheeler_wray_2022}
\begin{equation}
    b(s^\prime) = O(o \mid a, s^\prime) \int_{s} T(s^\prime \mid s, a) b(s) ds
\end{equation}
When the transition function, observation model, or belief representation is too complex to do an analytical belief update, we rely on approximate methods such as particle filters. 

The basic particle filter algorithm is shown in \cref{alg:particlefilter}. The belief is represented by a number of \emph{particles} which are samples of the state. Each update of the belief includes simulating the transition function for each particle to get a sample of the next state, computing the observation likelihood and then resampling according to the likelihood weights. The challenge with basic particle filtering is \emph{particle depletion} where the diversity of particles is reduced due to many of the particle having low likelihood weights. The problem is exacerbated with the number of belief updates and can require an intractably large number of initial state samples. 

An improvement to basic particle filtering, known as sequential importance resampling (SIR), iteratively adapts a proposal distribution to be closer to the posterior. The technique is outlined in \cref{alg:sirparticlefilter}. States are sampled from a proposal distribution that is initialized to the current belief. During each iteration samples are drawn from the proposal and weighted by the observation model and the importance weight. The proposal distribution is refit to these weighted samples and used in the next iteration, with the goal of being closer to the posterior distribution and producing better samples. Finally, a resampling step produces the final belief. 

\begin{algorithm}
\caption{Basic particle filter for belief $b$ (with $N$ particles), after action $a$ and observation $o$} \label{alg:particlefilter}
\begin{algorithmic}[1]
    \Function{BasicParticleFilterUpdate}{$b$, $N$, $a$, $o$}
    \State Sample transitions $s^\prime_i \sim T(\cdot \mid s_i, a)$ where $b = \{s_i\}_{i=1}^N$
    \State Compute likelihood weights $w_i \gets O(o \mid a, s^\prime_i)$
    \State $b^\prime \gets $ Resample $N$ particles from $\{s_i^\prime\}_{i=1}^N$ with weights $\{w_i\}_{i=1}^N$
    \State \textbf{return} $b^\prime$
    \EndFunction
\end{algorithmic}
\end{algorithm}

Our implementation of SIR makes several design decisions that were crucial to good performance. First, the proposal distribution is a kernel density estimate over the parameters that define the top surface geometry. Second, we incorporate all prior observations (not just the most recent) into the particle weight to ensure a good match with the observed data. Third, if there is insufficient particle diversity when fitting the next proposal, we fit to the \emph{elite samples} defined by the top $30$th percentile of samples in terms of their weights. Fourth, at each iteration we include 50\% samples from the prior distribution to mitigate the effect of poor proposals. Lastly, we run SIR until there are at least $N/2$ unique particles when resampling or we reach the pre-specified computational budget for the belief update.

\begin{algorithm}
\caption{Sequential Importance Resampling (SIR) particle filter for belief $b$ (with $N$ particles), after action $a$ and observation $o$, taking $N_k$ samples per iteration.} \label{alg:sirparticlefilter}
\begin{algorithmic}[1]
    \Function{SIRParticleFilterUpdate}{$b$, $N$, $a$, $o$}
    \State $q_1(s) \gets b(s)$
    \State $\mathcal{S} \gets \emptyset$
    \State $\mathcal{W} \gets \emptyset$
    \For{$k \in 1 \ldots K$}
        \State Sample $N_k$ states from proposal $s_i \sim q_k$ 
        \State Compute importance weight $w_i \gets b(s) / q_k(s)$
        \State Sample transitions $s^\prime_i \sim T(\cdot \mid s_i, a)$
        \State Include likelihood weight $w_i \gets w_i \cdot O(o \mid a, s^\prime_i)$
        \State Store particles $\mathcal{S} \gets \mathcal{S} \bigcup s_i$
        \State Store weights $\mathcal{W} \gets \mathcal{W} \bigcup w_i$
        \State $q_k \gets $ fit $\{s^\prime_i\}_{i=1}^{N_k}$ with weights $\{w_i\}_{i=1}^{N_k}$
    \EndFor{}
    \State $b^\prime \gets $ Resample $N$ particles from $\mathcal{S}$ with weights $\mathcal{W}$
    \State \textbf{return} $b^\prime$
    \EndFunction
\end{algorithmic}
\end{algorithm}

\begin{figure}
    \centering
    \includegraphics[width=\textwidth]{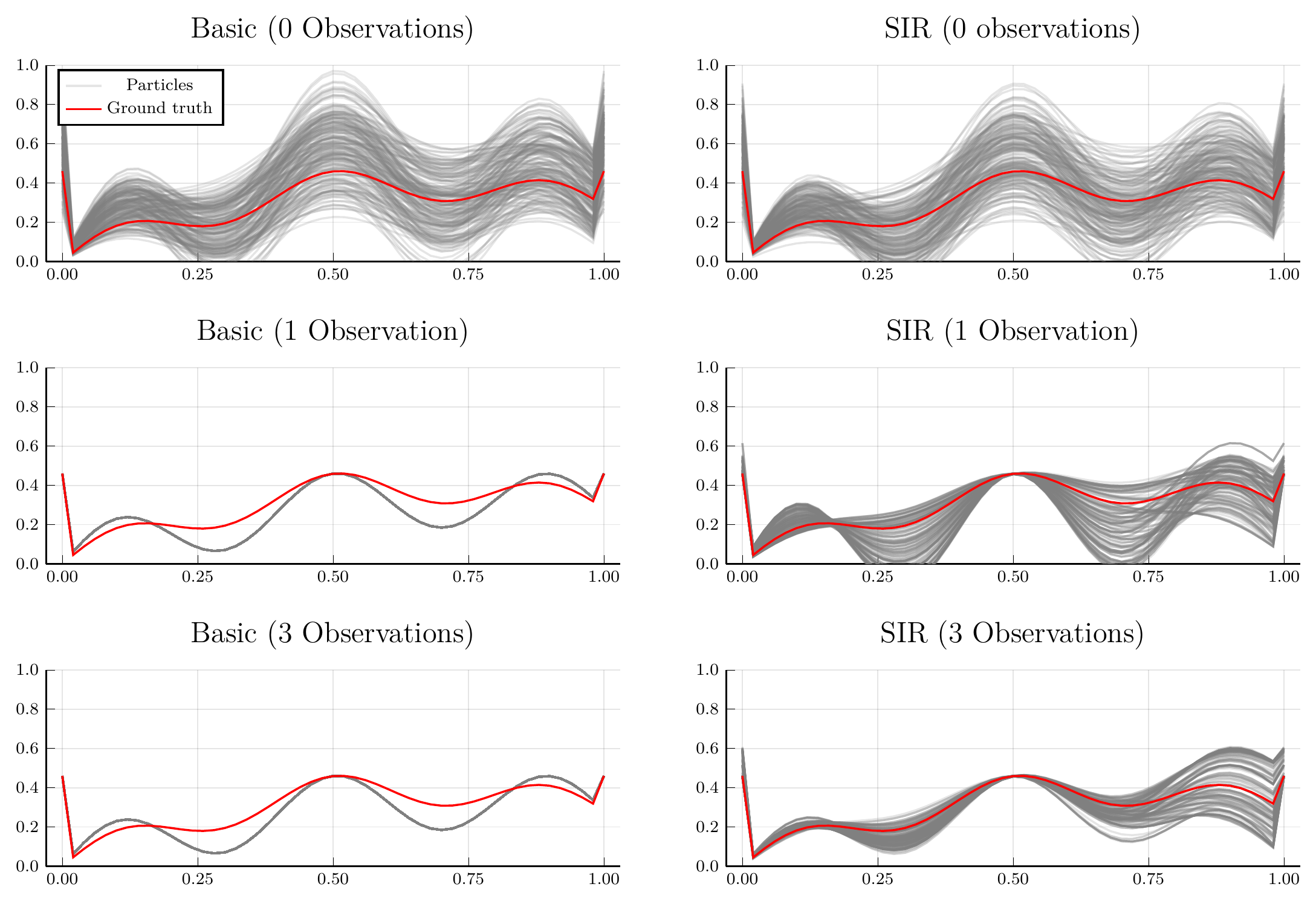}
    \caption{Comparison between basic particle filtering (left column) and SIR (right column) after different numbers of observations.}
    \label{fig:basic_SIR_comparison}
\end{figure}

Comparison between the basic particle filtering approach and SIR is shown in \cref{fig:basic_SIR_comparison}. For the same number of particles representing the belief, SIR leads to a better estimate of the posterior distribution while the basic particle filter suffers from particle depletion.

\section{Experiment Hyperparameters}
\label{sec:hyperparameters}
The parameters for the POMDP formulation and the hyperparameters for the experimental setup are included in \cref{tab:hyperparameters}.
\begin{table*}
    \caption{POMDP and Solver Hyperparameters\label{tab:hyperparameters}}
    \centering
    \begin{tabular}{@{}ll@{}} 
    \toprule
    Variable Description & Value \\
    \midrule
    Discount Factor ($\gamma$) & \num{0.9} \\
    $\lambda_{\rm leaked}$ & \num{-1000.0}\\
    $\lambda_{\rm exited}$ & \num{-1000.0}\\
    $\lambda_{\rm trapped}$ & \num{100.0}\\
    $\lambda_{\rm obs}$ & \num{-0.1}\\
    \ch{CO2} height std. dev. ($\sigma_{\rm height}$) & \num{0.01}\\
    \ch{CO2} thickness std. dev. ($\sigma_{\rm sat}$) & \num{0.01}\\
    Observation configurations & $\{ 0.25\!:\!0.25\!:\!0.75, 0.125\!:\!0.125\!:\!0.875\}$\\
    Injection rates & $\{ 0.01, 0.07\}$ \\
    Drill Locations & $\{ 0.1\!:\!0.1\!:\!0.9 \}$\\
    Distribution of $h_{\rm lobe}$ & $\mathcal{U}(0.05, 0.25)$\\
    Distribution of $h_{\rm center}$ & $\mathcal{U}(0.05, 0.5)$\\
    Distribution of $h_{\rm elev}$ & $\mathcal{U}(0.05, 0.5)$\\
    Distribution of normalized porosity ($\rho$) & $\mathcal{U}(0.5, 1.5)$\\
    \midrule
    Exploration Coefficient for UCB & \num{20.0} \\
    Observation widening exponent ($\alpha_{\rm obs})$ & \num{0.3} \\
    Observation widening coefficient ($k_{\rm obs})$ & \num{10.0} \\
    Action widening exponent ($\alpha_{\rm act}$) & \num{0.5} \\
    Action widening coefficient ($k_{\rm act})$ & \num{10.0} \\
    Number of tree queries ($N_{\rm query}$) & \num{5000}\\
    Estimation value for leaf nodes & $0.1 \times $ optimal return\\
    Particles for Basic Particle filter & \num{2000} $
    $ \\
    Particles for SIR ($N$) & \num{200} \\
    Samples per iteration for SIR & \num{100} \\
    Max CPU time for belief update & \num{60} seconds\\
    \bottomrule
    \end{tabular}
\end{table*}

\end{document}